\documentclass[letterpaper, 10 pt, twocolumn]{IEEEtran}

\usepackage[left=0.75in,right=0.75in,top=0.75in,bottom=0.8in]{geometry}

\usepackage[utf8]{inputenc}
\usepackage{graphics}
\usepackage{amsmath}
\usepackage{amssymb}
\usepackage{epsfig,wrapfig}
\usepackage{color}

\usepackage{tikz}
\usepackage{textcomp}
\usetikzlibrary{shapes,arrows}

\newcommand{\R}{\mathbb{R}}

\DeclareMathOperator{\im}{Im}

\DeclareMathOperator*{\argmax}{arg\,max}

\newtheorem{theorem}{Theorem}

\tikzset{%
  block/.style    = {draw, thick, rectangle, minimum height = 2em, minimum width = 2em},
  sum/.style      = {draw, circle, node distance = 1.5cm}, 
  input/.style    = {coordinate}, 
  output/.style   = {coordinate} 
}

\newcommand{\summ}{\Large$+$}

\newif\iflong
\longtrue

\title{\vspace*{0.25in}Distinguishing Aerial Intruders from Trajectory Data: A Model-Based Hypothesis-Testing Approach}

\author{
    David Petrizze, Kasra Koorehdavoudi, Mengran Xue, and Sandip Roy
    \thanks{
        The authors are with the School of Electrical Engineering and Computer Science at Washington State University. The authors acknowledge support from the Joint Center for Aerospace Technology Innovation and the United States National Science Foundation.
    }
}

\begin{document}
\maketitle
\thispagestyle{empty}

\begin{abstract}
    Motivated by security needs in unmanned aerial system (UAS) operations, an algorithm for identifying airspace intruders (e.g., birds vs. drones) is developed.  The algorithm is structured to use sensed intruder velocity data from Internet-of-Things platforms together with limited knowledge of physical models.  The identification problem is posed as a statistical hypothesis testing or detection problem, wherein inertial feedback-controlled objects subject to stochastic actuation must be distinguished by speed data.  The maximum {\em a posteriori} probability detector is obtained, and then is simplified to an explicit computation based on two points in the sample autocorrelation of the data.  The simplified form allows computationally-friendly implementation of the algorithm, and simplified learning from archived data.  Also, the total probability of error of the detector is computed and characterized.  Simulations based on synthesized data are presented to illustrate and supplement the formal analyses. 
\end{abstract}

\section{Introduction}

Unmanned aerial systems (UAS) support diverse applications ranging from emergency response to package delivery and infrastructure monitoring \cite{valvanis}. The growing use of UASs (drones) is bringing forth significant concerns with regard to security and privacy, as evidenced by numerous high-profile disruption events \cite{uavsecurity}. One particular concern is that the airspace of secure facilities, such as airports, prisons, ports, or power plants, may be violated by drones, which then may engage in destructive acts or extract sensitive information \cite{counteruas}. To address this concern, a range of technologies for detecting, classifying, and counteracting drone intrusions -- known collectively as CounterUAS technologies -- have been proposed.  Although some of these technologies show promise, they have not been adopted in operational systems at a wide-scale.  Key barriers to adoption include: 1) the difficulty of developing technologies that differentiate drones from other harmless intruders (e.g. birds) with high sensitivity and specificity; and 2) the high cost associated with deployment and persistent operation of many CounterUAS technologies.

New sensors deployed on Internet-of-Things (IoT) platforms hold great promise for enabling practical CounterUAS solutions.  In particular, new sensing systems -- such as small meta-materials-based radar systems -- can permit relatively high-frequency sensing of UAS trajectories at a reasonable cost, although typically at lower resolutions than traditional systems.  Meanwhile, IoT devices can allow distributed operation of these sensing technologies in the field at a reasonable cost.  However, use of these IoT solutions for CounterUAS requires development and computationally-efficient algorithms for detection and identification of threats.

Our goal  here is to explore algorithm development and deployment for IoT-based CounterUAS solutions, focusing particularly on the problem of distinguishing airspace or aerial intruders (e.g. birds vs. drones, or drones of different types) using trajectory data.  The intuition underlying our algorithms is that many intruders, including both drones and birds, are inertial objects which seek to follow piecewise-constant velocity trajectories using feedback in the presence of ambient stochastic disturbances (e.g. wind). However, they differ in mass, and also in the strengths of the applied feedback (e.g., drones typically have more tightly regulated velocities than birds). In consequence, the ambient velocity responses of airspace intruders to the ambient disturbances are varied, which intrinsically allows differentiation.  

Based on this intuition, we pose the problem of distinguishing airspace intruders of different types as a hypothesis testing or detection problem \cite{papoulis}, where the data generated under each hypothesis is governed by a feedback-controlled dynamical model with different parameters. We then use the standard machinery for hypothesis testing to formally construct an optimal algorithm for differentiation. The main contribution of the study is to then use the special structure of the hypothesis testing problem to gain insight into the optimal detector, and permit computationally-attractive learning and deployment of the detector.  Specifically, we demonstrate that the optimal detector reduces to a test on two points on the autocorrelation of the velocity; this form allows for scalable and easily deployable solutions on IoT platforms. Additionally, we show that the special problem structure allows for the computation of detection error probabilities, which serve as confidence levels on detection outcomes, and give insight into the data horizon needed for accurate differentiation of intruders.

From a methodological standpoint, the paper develops a way to distinguish feedback-controlled devices operating in noisy environments from sampled ambient time-series data.  Specifically, our work shows how measured data  can be used in tandem with limited physical model knowledge to enable improved detection, computationally-friendly detector implementation, and error analysis of the detector. 

The paper is organized as follows. The airspace intruder detector problem is presented in Section II. The optimal detector is developed in Section III, and then simplified to an explicit form  in Section IV.  The detection error is characterized in Section V. Finally, simulation examples are presented in Section VI.
\iflong
Note that two minor typos are present in the conference publication \cite{standard-version}; (\ref{eqn:total-error}) and Table \ref{tab:simparams} have since been corrected.
\fi

\section{Modeling and Problem Formulation}

We pose the problem of distinguishing airspace intruders as a statistical hypothesis testing or detection problem, wherein trajectory data together with simplified model knowledge is used for detection. 

\begin{figure}[htbp]
    \includegraphics[width=\linewidth]{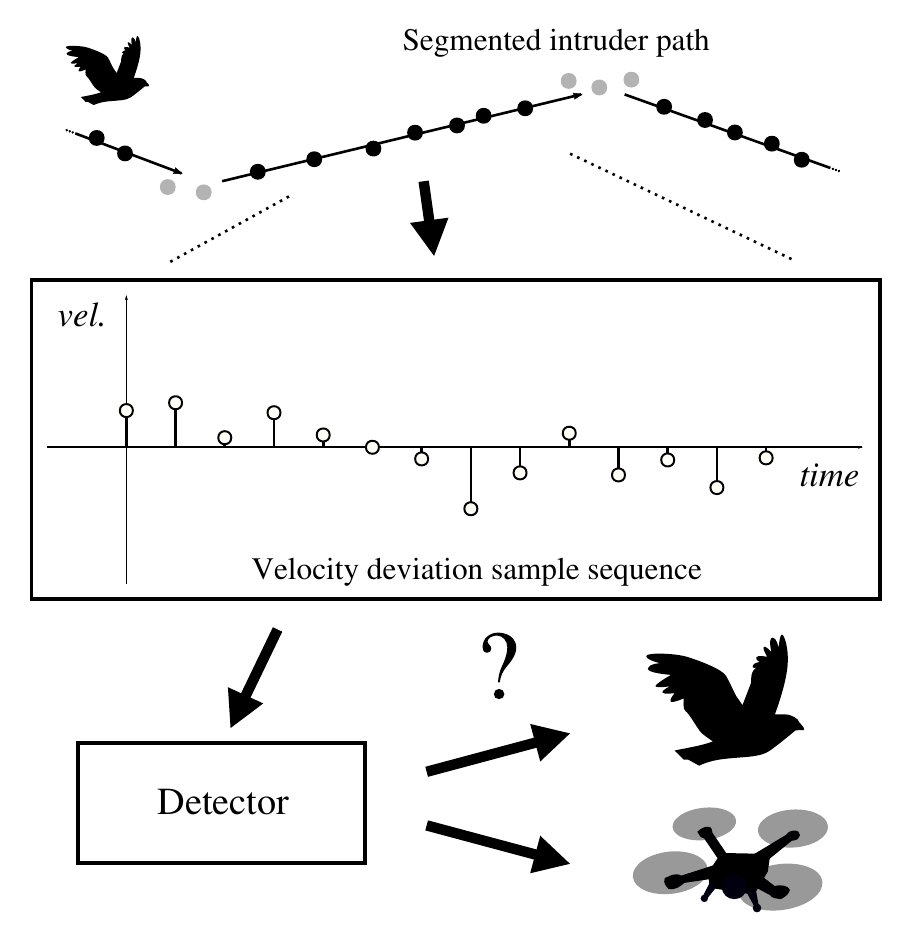}
    \caption{Overview of the problem formulation.}
    \label{fig:trajectories}
\end{figure}

\label{section:intruder-model}

Broadly, aerial intruders can be viewed as traversing a sequence of distinct flight segments while moving through the airspace (see Fig. \ref{fig:trajectories}). During each flight segment, the intruder can be modeled as seeking to follow a constant-velocity trajectory toward a waypoint, via application of a feedback control.  Once the waypoint is reached, the intruder then continues along a different flight segment, with a different velocity (speed, direction) goal. 

For this study, we assume that a sensing system (e.g., radar) obtains samples of the velocity profile, over a portion of the flight. Many recently developed low-cost sensing technologies, such as new lightweight radars and imaging technologies, are able to obtain relatively high frequency but low resolution data of this sort. In general, the measured profile  will include data from one or more constant-velocity flight segments of the aerial intruder. Here, we consider the circumstance that  data from only a single flight segment is available. Specifically, we consider using the deviation of the intruder's speed from the mean speed in its direction of flight, which approximates the deviation from the reference or goal speed, to distinguish the intruder.

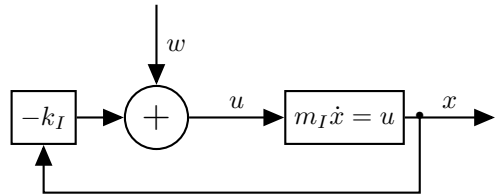
\begin{figure}
    \centering
    \begin{tikzpicture}[auto, thick, node distance=1.5cm, >=triangle 45]
        \draw
        node at (4, 0) [block                      ] (sysblock)   {$m_I \dot{x} = u$}
        node at (6, 0) [output                     ] (output_x)   {}
        node           [block                      ] (kblock)     {$-k_I$}
        node           [sum,    right of=kblock    ] (sum2)       {\summ}
        node           [input,  above of=sum2      ] (input_w)    {};
        \draw[->](kblock)     -- node {}           (sum2);
        \draw[->](input_w)    -- node {$w$}        (sum2);
        \draw[->](sum2)       -- node {$u$}        (sysblock);
        \draw[->](sysblock)   -- node {$x$}        (output_x);
        \draw[->](5,0)        -- (5,-1) -| node {} (kblock);
        \draw node at (5,0) {\textbullet};
    \end{tikzpicture}
    \caption{The feedback control model representing an intruder's speed deviation $x$ from the
    reference speed is illustrated.  The mass and strength of the feedback gain are modeled as
    differing for intruders of different types; for instance, drones typically tightly regulate
    their speed (i.e. use a bigger gain $k_I$) as compared to birds.}
    \label{fig:object-feedback}
\end{figure}

While a detector could be built by only considering measured speed data, incorporation of some model knowledge about intruder flight should allow for improved detection, and also perhaps allow more computationally friendly detector learning and use.  With this in mind,  a simplified mathematical model for the deviation of the intruder's speed from the reference speed is considered. To permit an easier development, let us assume that there are two types of intruders indicated by the variable $I$, say $I=1$ (e.g. bird) and $I=2$ (e.g. drone). The intruder's speed deviation from the reference is denoted as $x(t)$, and is modeled as arising from a feedback-controlled inertial object which is subject to a zero-mean white Gaussian disturbance. Specifically, the deviation is modeled as:
\begin{equation}
    \label{eqn:intruder-model}
    m_I \dot{x} = -k_I x + w \textrm{,}
\end{equation}
where $I$ indicates the type of intruder, $m_I$ is the intruder's mass, $k_I$ is a lumped feedback term which indicates the strength of the regulatory effort imposed by the intruder's velocity control systems, and $w$ is a force disturbance (e.g. an environmental force) which is modeled as a zero-mean white-noise signal with intensity $q$ (i.e. the noise autocorrelation is $R_{ww}(t)=q \delta(t)$). This model is illustrated in Fig. \ref{fig:object-feedback}. 

A sensor (e.g. a radar system) is assumed to make measurements of an intruder's velocity profile at sample times, from which the velocity deviation at these times can be computed. Thus, we assume that the measurement signal
\begin{equation}
    \label{eqn:sampling-relationship}
    y[k] = x(kT)
\end{equation}
is available at times $k=0,\hdots, k_f-1$, where $T$ is the sampling interval, $k=0$ corresponds to the first available velocity measurement, and $k_f$ measurements are available in total.

Our goal is to develop a method for distinguishing the type of an intruder, using the measurement signal $y[k]$. We formulate the problem of distinguishing the intruder type as {\em maximum a posteriori probability (MAP) detection} or {\em hypothesis testing} problem, wherein the most likely intruder type given the observation sequence is selected as the intruder type.   Mathematically, the MAP detector $\widehat{I}$ is given by:
\begin{equation}
    \widehat{I}=\argmax_{i=1,2} \Pr(I=i \, |\, y[0],\hdots, y[k_f-1]).
\end{equation}
We note that computation of the MAP detector requires prior probabilities of the intruder types, $\Pr(I=1)=p$ and $\Pr(I=2)=1-\Pr(I=1)=1-p$. Alternately, absent priors, a varying threshold can be used to distinguish between the intruder types to tune sensitivity vs specificity; see \cite{papoulis}.   We also seek to compute probability of error metrics for the detector, which are valuable both for providing real-time confidence intervals, and for overall performance evaluation of the detector.

Finally, it is noted that the model for an intruder's speed deviation is a gross over-simplification, in that it does not capture any multi-time-constant dynamics of the intruder, nonlinearities such as wind drag, and structure in the spectrum of the disturbance.  We deliberately use an over-simplified model because of the difficulty inherent to building and parameterizing detailed models of intruder dynamics.  Thus, our approach is to use a condensed model of the physics -- which captures the salient constraints on the physical dynamics resulting from inertia and regulation -- together with data to get simple and hopefully robust rubrics for such a detector.  Our analysis will show that the abstracted model yields a very simple sufficient statistic for detection, which is promising for differentiating airspace intruders in practical settings.

\section{Detector}
\label{sec:our-detector}

A formal expression for the MAP detector for distinguishing airspace intruders is developed in the following theorem, using the standard methodology for hypothesis testing. To present the result, we use the notation ${\bf y}$ for the vector of observations used for detection, i.e. ${\bf y} = \bigl( y[0], y[1], \cdots, y[k_f-1] \bigr)^T$.

\begin{theorem}
    \label{thm:detector}
    The MAP detector for the intruder-identification problem is given by:
    \begin{equation}
        \label{eqn:basic-detector}
        \widehat{I}({\bf y}) = \begin{cases}
            1 & \textrm{ for } {\bf y}^T Q {\bf y} \leq z \\
            2 & \textrm{ otherwise }
        \end{cases}
        \textrm{.}
    \end{equation}
    Here, $Q$ is given by:
    \begin{equation}
        \label{eqn:Q}
        Q = \Sigma_1^{-1}-\Sigma_2^{-1}
        \textrm{,}
    \end{equation}
    where $\Sigma_I$ is the covariance matrix governing the velocity observations of intruder type
    $I$. The $i,j$-th element of $\Sigma_I$ is:
    \begin{equation}
        \label{eqn:sigma-definition}
        \left(\Sigma_I\right)_{ij} = 
        \frac{q}{2 k_I m_I}e^{-\frac{k_I}{m_I}T\left|i-j\right|} =
        \alpha_I\rho_I^{\left|i-j\right|}
        \textrm{,}
    \end{equation}
    where $T$ is the sampling period, $q$ is the wind noise magnitude, $m_I$ is the intruder's
    effective mass, $k_I$ is the intruder's effective feedback gain, and
    $\rho_I=e^{-\frac{k_I}{m_I}T}$. Finally, the detector threshold $z$ is given by:
    \begin{equation}
        \label{eqn:z}
        z = \ln\left({\frac{\Pr\left(I=1\right)^2\left|\Sigma_2\right|}{\Pr\left(I=2\right)^2\left|\Sigma_1\right|}}\right)
        \textrm{.}
    \end{equation}
\end{theorem}

\iflong
    \begin{IEEEproof}
        The observation sequence under each hypothesis is a zero mean Gaussian process, with different autocorrelation functions.  The form of the hypothesis test derives readily from the joint Gaussian distribution of the observations under each hypothesis (see e.g. \cite{papoulis}).  The covariance matrices can then be obtained from the statistical analysis of a first-order system driven by white noise.
    
        Specifically, to derive the form of the MAP detector, the joint probability density function for the measurement sequence under each hypothesis (intruder type) is compared. Based on the Gaussian density function, the MAP detector specifies $\widehat{I}=1$ if the following expression holds:
        \begin{equation}
            \label{eqn:raw-detector}
            \frac{\Pr\left(I=1\right)}{\sqrt{\left|2\pi\Sigma_1\right|}}e^{-\frac{1}{2}{\bf y}^T\Sigma_1^{-1}{\bf y}} > \frac{\Pr\left(I=2\right)}{\sqrt{\left|2\pi\Sigma_2\right|}}e^{-\frac{1}{2}{\bf y}^T\Sigma_2^{-1}{\bf y}},
        \end{equation}
        and specifies $\widehat{I}=2$ otherwise. This expression readily simplifies to give
        (\ref{eqn:basic-detector}), (\ref{eqn:Q}), and (\ref{eqn:z}).
    
        The entries in the covariance matrix $\Sigma_{ij}$ follow from the statistical analysis of the dynamical model of the intruder (\ref{eqn:intruder-model}).  To characterize the covariance, we first note that the  transfer function from the white noise drive to the velocity signal is given by:
        \begin{equation}
            \label{eqn:object-transfer-function}
            \frac{X(s)}{W(s)} = \frac{1}{m_I s + k_I}
            \textrm{.}
        \end{equation}
        From the transfer function, it readily follows that the autocorrelation of the (stationary) velocity response is given by:
        \begin{equation}
            \label{eqn:object-autocorrelation}
            R_{vv}(\tau) = \frac{q}{2 k_I m_I} e^{-\frac{k_I}{m_I}\left|\tau\right|}
            \textrm{.}
        \end{equation}
        Considering samples as specified by (\ref{eqn:sampling-relationship}), it follows that the covariance matrix $\Sigma_I$ is given by (\ref{eqn:sigma-definition}).
    \end{IEEEproof}
\fi

\section{Simplified Computation of the Detector}

The expression for the detector given in Theorem 1 allows for an algorithmic computation of the MAP hypothesis.  However, the computation on its face is rather intensive, requiring inversion and computation of determinants for two dense covariance matrices whose dimensions grow with the number of samples. Further, the determinant and inverse calculations are prone to numerical instability unless a very high precision is used. Thus, these inversion and determinant computations may be computationally unattractive when they must be performed using low-cost hardware (e.g., a microcontroller).  On the other hand, because CounterUAS technologies would typically be used in outdoor spaces with harsh environments, persistent transmission of data to a central server is also unappealing from an energy-use standpoint, and also may be subject to delays and errors.  Given these concerns, there is a significant motivation to pursue a simplified, explicit computation of the MAP detector.  

In addition to these computational challenges, the expression for the MAP detector in Theorem 1 also does not give intuition into how the measured data is processed for detection -- the detector uses an apparently arbitrary quadratic form of the data. This lack of insight is also problematic for field implementation of the technology for two reasons.  First, the expression does not identify whether a small set of sufficient statistics can be used for detection; if such a set of sufficient statistics could be determined, this would simplify detector computation and also possibly allow for simplified learning of the detector if the model was unknown.  By the same token, insight into the detector form may allow for more robust detection algorithms when model parameters are partially unknown or subject to error.

With these motivations, the interesting structure of $\Sigma_I$ (\ref{eqn:sigma-definition}) is  particularly appealing from a simplification standpoint. The following theorem demonstrates that the matrix inverse and determinant can be computed exactly, thus yielding a simple, explicit expression for the MAP detector.

\begin{theorem}
    The MAP detector for the intruder-identification problem is given by:
    \begin{equation}
        \label{eqn:reduced-detector}
        \widehat{I}({\bf y}) = \begin{cases}
            1, &
                \begin{aligned}
                    & a k_f R_{yy}(0) + b k_f R_{yy}(1) + \\
                    & c \left({\bf y}_0^2 + {\bf y}_{k_f-1}^2\right)
                \end{aligned}
                \leq z \\
            2, & \textrm{otherwise}
        \end{cases}
        \textrm{,}
    \end{equation}
    where $R_{yy}(k)$ denotes the sample autocorrelation of the elements of ${\bf y}$. The constants
    $a$, $b$, and $c$ are given as follows:
    \begin{align}
        \label{eqn:reduced-detector-a}
        a &= \frac{1 + \rho_1^2}{\alpha_1\left(1 - \rho_1^2\right)} -
              \frac{1 + \rho_2^2}{\alpha_2\left(1 - \rho_2^2\right)}
        \textrm{,} \\
        \label{eqn:reduced-detector-b}
        b &= \frac{2\rho_2}{\alpha_2\left(1 - \rho_2^2\right)} -
              \frac{2\rho_1}{\alpha_1\left(1 - \rho_1^2\right)}
        \textrm{, and} \\
        \label{eqn:reduced-detector-c}
        c &= \frac{\rho_2^2}{\alpha_2\left(1 - \rho_2^2\right)} -
              \frac{\rho_1^2}{\alpha_1\left(1 - \rho_1^2\right)}
        \textrm{.}
    \end{align}
    Similarly, $z$ is given by:
    \begin{equation}
        \label{eqn:reduced-z}
        \begin{split}
            z &= 2\ln\left(\frac{\Pr(I=1)}{\Pr(I=2)}\right) \\
              &+ k_f\ln\left(\frac{\alpha_2}{\alpha_1}\right) \\
              &+ (k_f-1)\ln\left(\frac{1 - \rho_2^2}{1 - \rho_1^2}\right)
            \textrm{.}
        \end{split}
    \end{equation}
\end{theorem}

\iflong
    \begin{IEEEproof}
        $\Sigma_I$ is readily identified as a Kac-Murdock-Szegö matrix \cite{kms-inverse}, with the
        following analytical inverse:
        \begin{equation}
            \label{eqn:sigma-inverse}
            \Sigma_I^{-1} = \frac{1}{\alpha_I\left(1 - \rho_I^2\right)}
            \begin{bmatrix}
                1       & -\rho_I       & \dots  & 0            & 0        \\
                -\rho_I & 1 + \rho_I^2  & \dots  & 0            & 0        \\
                \vdots  & \vdots        & \ddots & \vdots       & \vdots   \\
                0       & 0             & \dots  & 1 + \rho_I^2 & -\rho_I  \\
                0       & 0             & \dots  & -\rho_I      & 1        \\
            \end{bmatrix}
            \textrm{,}
        \end{equation}
        and a determinant given by:
        \begin{equation}
            \label{eqn:sigma-determinant}
            \left|\Sigma_I\right| = \alpha_I^{k_f}(1 - \rho_I^2)^{k_f - 1}
            \textrm{.}
        \end{equation}
        From here, (\ref{eqn:reduced-detector}), (\ref{eqn:reduced-detector-a}),
        (\ref{eqn:reduced-detector-b}), and (\ref{eqn:reduced-detector-c}) follow by evaluating the
        detector expression: ${\bf y}^T \left(\Sigma_1^{-1} - \Sigma_2^{-1}\right) {\bf y}$, and
        simplifying the result. The expression for $z$ in (\ref{eqn:reduced-z}) follows in the same way.
    \end{IEEEproof}
\fi

The explicit expression for the detector clarifies the sufficient statistic needed for detection, and gives intuition into how aerial intruders can be distinguished.  Specifically, the detector requires only two points in the sample autocorrelation: 1) the sample variance of the measurement signal ${\bf y}$, and 2) the covariance between subsequent samples in the measurement signal (i.e., measurements with one time step delay between them), provided that the intruder's persistent trajectory is measured over a sufficiently long horizon.  (The detector also includes an "edge effect" which reflects measurement limits at the initiation and end of measurement, but these terms become negligible as $k_f$ increases.) Meanwhile, the threshold for detection, $z$, is determined by the masses and control gains of the intruders, together with the sampling interval and the {\em a priori} probabilities of each intruder type.

Noting that the detector threshold can be modified to trade off sensitivity and specificity, detection essentially depends only on calculation of the variance and a single-shift covariance of the data. The simple detector form reflects the fact that intruders are modeled as an inertial mass with a proportional feedback.  Provided that intruders can be roughly modeled in this way, i.e. they have a dominant response time constant, then this simple detector may be expected to allow for effective differentiation from ambient data. 

Importantly, a detector of this form can be learned without knowledge of system parameters and given inaccuracies in the model form, using training data.  In particular, noting that the detector is based on a linear combination of two points in the autocorrelation, a detector can be built by computing these two points in archived measurements of the two intruder types.  These statistics can then be presented on a scatter plot to identify the detection rule.  Thus, an entirely data-theoretic implementation of the detector is possible, even though the approach takes advantage of the dynamical model structure or constraints of aerial intruders (i.e., inertia, presence of feedback).

\section{Detector Error}

We next seek to characterize the performance of the MAP detector, as measured by its probability of error. The error-probability calculation provides a means for giving confidence intervals around the detection goal, which indicates the fidelity of the estimate to users of the detector.  The presentation of confidence intervals is important for field implementation of the technology, because manual intervention to resolve threats is often costly, and hence operations would benefit from an indication of detector fidelity prior to intervention. An analysis of the detector's performance also gives insight into the amount of data required for differentiation of aerial intruders, and hence allows understanding of whether detection from ambient data is practical or not.  Intuitively, the gap between the two covariance matrices $\Sigma_1$ and $\Sigma_2$, as well as the number of velocity samples $k_f$, should determine the probability of correctly distinguishing the intruder.

Per standard hypothesis-testing methodologies, two types of error probabilities may be of interest. First, the total {\em a priori} probability of error of the detector, i.e. $\Pr(\widehat{I}({\bf y}) \neq I)$, is of interest as an overall fidelity metric for the detector. We denote this total error probability by $\Pr(E_T)$, where $E_T$ is the event that a misdetection occurs ($\widehat{I}({\bf y}) \neq I$).  The total error probability is an indication of the effectiveness of the detector over repeated trials, and is a function of only the system's parameters.

Second, the probability of mis-detection given a sequence of measurements, i.e. $\Pr(\widehat{I}({\bf y}) \neq I \, | \, {\bf y})$, is of interest as a real-time confidence interval for a deployed detector. This conditional error probability can readily be computed in terms of the prior probabilities and the Gaussian observation-sequence probabilities, via an application of Baye's rule.
 
\subsection{A Priori Detector Error}

The total error probability does not allow for a closed-form expression.  However, using analyses of quadratic forms of Gaussian random variables, a convenient form for numerical computation of the error can be derived.  In addition, by then exploiting the specific system structure for our detection problem, the factors influencing the detection probability also can be characterized, and some structural insight can be gained into the detector performance.  The expression that allows for numerical computation of the total error probability is presented in the following theorem.

\begin{theorem}
    The total probability of error for the detector can be computed as:
    \begin{equation}
        \label{eqn:total-error}
        \Pr \left( E_T \right) = \begin{aligned}
            & \Pr \left(I = 2 \right) F_Z \left( Q, \Sigma_2, z \right) + \\
            & \Pr \left(I = 1 \right) \left[ 1 - F_Z \left( Q, \Sigma_1, z \right) \right]
        \end{aligned}
        \textrm{,}
    \end{equation}
    where $Q$, $\Sigma_1$, and $\Sigma_2$, and $z$ are defined in the presentation of the detector (Theorems 1 and 2). The term $F_Z (Q, \Sigma, z)$, a numerically-computed inverse Fourier transform, is given by:
    \begin{equation}
        \label{eqn:F_Z}
        F_Z(Q, \Sigma, z) =
            \frac{1}{2} - \sum_{i=0}^N \frac
                { \im \left[ \phi \left( \Delta ( i + \frac{1}{2} ) \right) e^{-j z \Delta \left( i + \frac{1}{2} \right) } \right] }
                { \pi \left( i + \frac{1}{2} \right) }
        \textrm{,}
    \end{equation}
    where $j$ is the imaginary constant, and $\phi(\omega)$ is defined as:
    \begin{equation}
        \label{eqn:z-characteristic}
        \phi(\omega) = \prod_{k=1}^{k_f} \left( 1 - 2 j \omega \lambda_k \right)^{-\frac{1}{2}}
        \textrm{.}
    \end{equation}
    Here, $\lambda_k$ is the $k$th eigenvalue of the matrix product $Q\Sigma$.
    
\end{theorem}

\iflong
    \begin{IEEEproof}
        Assume that ${\bf y}$ contains $k_f$ velocity samples with zero mean, and that $Q$, $z$, $\Sigma_1$, and $\Sigma_2$ characterize a detector as defined by Section \ref{sec:our-detector}. Intuitively, the probability that the detector concludes falsely is given by:
        \begin{align}
            \nonumber
            \Pr \left( E_1 \right) &= \Pr({\bf y}^T Q {\bf y} \leq z \mathop{\cap} I = 2)
            \textrm{, and} \\
            \nonumber
            \Pr \left( E_2 \right) &= \Pr({\bf y}^T Q {\bf y} > z \mathop{\cap} I = 1)
            \textrm{.}
        \end{align}
        Here, $\Pr(E_1)$ and $\Pr(E_2)$ denote the probability of the detector misidentifying intruder types 1 and 2, respectively. Clearly, $\Pr(E_T) = \Pr(E_1) + \Pr(E_2)$, and through simple manipulation, we can write $\Pr(E_T)$ in terms of $\Pr({\bf y}^T Q {\bf y} \leq z \mid I = ...)$.
        \begin{equation}
            \nonumber
            \Pr \left( E_T \right) = \begin{aligned}
                & \Pr(I=2)\Pr({\bf y}^T Q {\bf y} \leq z \mid I = 2) + \\
                & \Pr(I=1)[1 - \Pr({\bf y}^T Q {\bf y} \leq z \mid I = 1) ]
            \end{aligned}
        \end{equation}
        We can interpret the above conditional probability expression in terms of the CDF of the random variable $Z$, where $Z$ = $Y^T Q Y$, and the random vector $Y$ is Gaussian distributed with covariance $\Sigma_I$. This yields (\ref{eqn:total-error}), and hence the notation $F_Z$.
    
        From \cite{linear-models-chi-squared-distribution}, $Z$ is observed to have a \textit{generalized chi-squared distribution}, with a characteristic function defined by (\ref{eqn:z-characteristic}). The result follows from application of \cite{davies-inverse}.
    \end{IEEEproof}
\fi

To ensure that (\ref{eqn:F_Z}) is accurate, the constants $N$ and $\Delta$ must be chosen appropriately to ensure that an accuracy threshold $E$ (where $0 < E < 1$) is met.  The threshold indicates the maximum deviation between the true and computed error due to truncation (selection of $N$) and resolution (selection of $\Delta$) inaccuracies.  To achieve this accuracy level, we claim that it is sufficient to assign
\begin{equation}
    \Delta = \frac{2 \pi t}{\ln(\theta) - \ln\left(\frac{E}{2}\right)}
    \textrm{, and}
\end{equation}
\begin{equation}
    N = \left\lceil
        \frac{1}{2\Delta \left| \lambda_{min} \right|} \left( \frac{\pi}{4} E k_f \right)^{-\frac{2}{k_f}}
    \right\rceil
    \textrm{,}
\end{equation}
where $\theta$ is defined as:
\begin{equation}
    \theta =
        \max \left\{
            e^{ zt} \prod_{k=1}^{k_f} \left(1+2t\lambda_k\right)^{-\frac{1}{2}},
            e^{-zt} \prod_{k=1}^{k_f} \left(1-2t\lambda_k\right)^{-\frac{1}{2}}
        \right\}
    \textrm{.}
\end{equation}
Here, $t$ is chosen such that $0 < t < \frac{1}{2 \left| \lambda_{max} \right|}$; $\lambda_{min}$ and $\lambda_{max}$ represent the $\lambda_k$ with minimum and maximum magnitude, respectively.

\iflong
    To justify the bound on $N$, we notice that the sum's truncation error, $E$, is given by:
    \begin{equation}
        E = \left|
            \sum_{k=N+1}^{\infty}
                \frac{1}{\pi(k+\frac{1}{2})} \Im \left[
                    \phi\left((k+\frac{1}{2})\Delta\right) e^{-j (k+\frac{1}{2})\Delta z}
                \right]
        \right|
        \textrm{.}
    \end{equation}
    By the triangle inequality, we know that:
    \begin{equation}
        E \leq
            \sum_{k=N+1}^{\infty}
                \frac{1}{\pi(k+\frac{1}{2})} \left| \Im \left[
                    \phi\left((k+\frac{1}{2})\Delta\right) e^{-j (k+\frac{1}{2})\Delta z}
                \right] \right|
        \textrm{.}
    \end{equation}
    Considering the maximum possible magnitude of the expression inside the absolute value, we similarly find:
    \begin{equation}
        E \leq
            \sum_{k=N+1}^{\infty}
                \frac{1}{\pi(k+\frac{1}{2})} \left|
                    \phi\left((k+\frac{1}{2})\Delta\right)
                \right|
        \textrm{.}
    \end{equation}
    Subsituting $\phi(u)$, we have:
    \begin{align}
        E &\leq
            \sum_{k=N+1}^{\infty}
                \frac{1}{\pi(k+\frac{1}{2})} \left|
                    \prod_{i=1}^{k_f} \left(1-2ju\lambda_i\right)^{-\frac{1}{2}}
                \right| \\
        &=
            \sum_{k=N+1}^{\infty}
                \frac{1}{\pi(k+\frac{1}{2})}
                    \prod_{i=1}^{k_f} \left| \left(1-2ju\lambda_i\right) \right| ^{-\frac{1}{2}}
        \textrm{,}
    \end{align}
    where $u = (k+\frac{1}{2})\Delta$. Note that in our case, $Q$ is symmetric, and thus each $\lambda_i$ is real. By ignoring the real component, the magnitude of the denominator in each term of the product will strictly decrease, meaning:
    \begin{align}
        E &\leq
            \sum_{k=N+1}^{\infty}
                \frac{1}{\pi(k+\frac{1}{2})}
                    \prod_{i=1}^{k_f} \left( 2 u \left| \lambda_i \right| \right) ^{-\frac{1}{2}}
        \textrm{.}
    \end{align}
    By reducing each $\lambda_i$ to the $\lambda_i$ with the smallest magnitude ($\lambda_{min}$), the denominator will again decrease, so we find:
    \begin{align}
        E &\leq
            \sum_{k=N+1}^{\infty}
                \frac{\left( 2 u \left| \lambda_{min} \right| \right) ^{-\frac{k_f}{2}}}{\pi(k+\frac{1}{2})}
        \textrm{.}
    \end{align}
    Substituting for $u$, we are left with:
    \begin{align}
        E &\leq
            \sum_{k=N+1}^{\infty}
                \frac{\left( 2 (k+\frac{1}{2})\Delta \left| \lambda_{min} \right| \right) ^{-\frac{k_f}{2}}}{\pi(k+\frac{1}{2})} \\
        &=
            \frac{1}{\pi} \left( 2 \Delta \left| \lambda_{min} \right| \right)^{-\frac{k_f}{2}}
            \sum_{k=N+1}^{\infty}
                \left( k+\frac{1}{2} \right) ^{-(\frac{k_f}{2} + 1)}
        \textrm{.}
    \end{align}
    We can yet again reduce the denominator in each term of the sum to find:
    \begin{align}
        E &\leq
            \frac{1}{\pi} \left( 2 \Delta \left| \lambda_{min} \right| \right)^{-\frac{k_f}{2}}
            \sum_{k=N+1}^{\infty}
                k ^{-(\frac{k_f}{2} + 1)}
        \textrm{,}
    \end{align}
    where the infinite sum is readily identified as a (convergent) p-series. We can bound this sum with the appropriate integral as follows:
    \begin{align}
        E &\leq
            \frac{1}{\pi} \left( 2 \Delta \left| \lambda_{min} \right| \right)^{-\frac{k_f}{2}}
            \int_{K}^{\infty}
                \tau ^{-(\frac{k_f}{2} + 1)} \mathop{d\tau} \\
        &=
            \frac{1}{\pi} \left( 2 \Delta \left| \lambda_{min} \right| \right)^{-\frac{k_f}{2}}
            \frac{2}{k_f} N ^{-\frac{k_f}{2}} \\
        &=
            \frac{2}{k_f\pi} \left( 2 N \Delta \left| \lambda_{min} \right| \right)^{-\frac{k_f}{2}}
            := E_{upper}
        \textrm{.}
    \end{align}
    Finally, we can solve for $N$ in terms of a given upper bound on the error, $E_{upper}$:
    \begin{equation}
        N = \left\lceil
            \frac{1}{2\Delta \left| \lambda_{min} \right|} \left( \frac{k_f \pi}{2} E_{upper} \right)^{-\frac{2}{k_f}}
        \right\rceil
    \end{equation}
    
    To justify the bound on $\Delta$, we work from the following expression for the resolution inaccuracy which is developed in \cite{davies-inverse}:
    \begin{equation}
        E = \max \left\{ \Pr\left(Z < z - \tfrac{2\pi}{\Delta}\right), \Pr\left(Z > z + \tfrac{2\pi}{\Delta}\right) \right\}
        \textrm{.}
    \end{equation}
    It turns out to be relatively straightfoward to find $\Delta > 0$ so as bound this quantity, through the use of the Chernoff bounds.
    
    We know that $Z$'s moment generating function, $M_Z(t)$, is strictly positive and real valued for $t \in (T_1, T_2)$, where the range $(T_1, T_2)$ contains the origin. Hence, the Chernoff bounds for each tail probability are given by:
    \begin{align}
        \Pr\left(Z < z - \tfrac{2\pi}{\Delta}\right) &\leq M_Z(s)e^{-s\left(z - \tfrac{2\pi}{\Delta}\right)} \textrm { for } T_1<s<0 \\
        \Pr\left(Z > z + \tfrac{2\pi}{\Delta}\right) &\leq M_Z(t)e^{-t\left(z + \tfrac{2\pi}{\Delta}\right)} \textrm { for } 0<t<T_2
        \textrm{.}
    \end{align}
    The quantity $E$ will thus be bounded by the respective bounds for each tail as follows:
    \begin{equation}
        E \leq \max \left\{
            M_Z(s)e^{-s\left(z - \tfrac{2\pi}{\Delta}\right)},
            M_Z(t)e^{-t\left(z + \tfrac{2\pi}{\Delta}\right)}
        \right\}
        \textrm{.}
    \end{equation}
    Substituting $M_Z(t)$, we find:
    \begin{equation}
        E \leq
             \max \left\{
                \begin{aligned}
                    e^{ \tfrac{2\pi}{\Delta}s} e^{-zs} \prod_{i=1}^{k_f} \left(1-2s\lambda_i\right)^{-\frac{1}{2}}, \\
                    e^{-\tfrac{2\pi}{\Delta}t} e^{-zt} \prod_{i=1}^{k_f} \left(1-2t\lambda_i\right)^{-\frac{1}{2}}
                \end{aligned}
            \right\}
        \textrm{.}
    \end{equation}
    From here, let $t = -s$, suitably chosen such that the condition on $M_Z(t)$ is satisfied. (For example, let $t = \frac{1}{4\left|\lambda_{max}\right|}$, where $\lambda_{max}$ is the $\lambda_i$ with the largest magnitude.) This substitution reduces our expression for $E$ to the following:
    \begin{equation}
        \label{eqn:P0-theta-eqn}
        E \leq \theta e^{-\frac{2 \pi t}{\Delta}} := E_{upper}
        \textrm{,}
    \end{equation}
    where $\theta$ is defined as:
    \begin{equation}
        \theta =
            \max \left\{
                e^{ zt} \prod_{i=1}^{k_f} \left(1+2t\lambda_i\right)^{-\frac{1}{2}},
                e^{-zt} \prod_{i=1}^{k_f} \left(1-2t\lambda_i\right)^{-\frac{1}{2}}
            \right\}
        \textrm{.}
    \end{equation}
    Once $\theta$ has been computed, equation \ref{eqn:P0-theta-eqn} can be used to compute $\Delta$ in terms of an upper bound on error, $E_{upper}$, as follows:
    \begin{equation}
        \Delta = \frac{2 \pi t}{\ln(\theta) - \ln(E_{upper})}
        \textrm{.}
    \end{equation}
\fi

As demonstrated via simulation in section \ref{sec:examples}, Theorem 3 provides an appealing method for computing the total probability of error of the airspace-intruder detector.  Also of importance, the theorem characterizes the factors that influence the total error probability.  Crucially, the theorem highlights that the detector error probability is dependent on the eigenvalues of the matrix products $Q\Sigma_1$ and $Q\Sigma_2$.  However, we notice that $Q\Sigma_1=(\Sigma_1^{-1}-\Sigma_2^{-1}) \Sigma_1 =I-Sigma_2^{-1} \Sigma_1$, and similarly $Q\Sigma_2=I-\Sigma_1^{-1}\Sigma_2$.  With some effort, it can be seen that the rows of $\Sigma_2^{-1} \Sigma_1$ are related to the deconvolution of the impulse response of the type 2 intruder from that of the type 1 intruder.  This matrix is close to the identity matrix if the two impulse responses are similar, which is the case if both the ratio and product of the control gain and mass are similar.  In this case, $Q\Sigma_1$ and $Q\Sigma_2$ are nearly $0$, and it follows immediately that the error probability is equal to the {\em a priori}.  Meanwhile, if the impulse responses differ significantly (equivalently, their spectra are more distinct), the error probability decreases quickly with the number of samples $k_f$.

\subsection{Detector Error Given Measurements}

In addition to knowing the total probability of error, it is useful to know the probability of error given some measured velocity sequence ${\bf y}$. This probability, as well as an expression for the likelihood ratio for the alternate hypothesis, is stated in the following theorem:

\begin{theorem}
    \label{thrm:error-given-y}
    For a sequence of velocity deviation measurements in ${\bf y}$, the probability that the detector has concluded incorrectly is given by:
    \begin{equation}
        \Pr \left( I \neq \widehat{I} \mid {\bf Y} = {\bf y} \right) = \left(1 + \frac{1}{r}\right)^{-1}
        \textrm{,}
    \end{equation}
    where $r$ is the probability ratio for the alternate hypothesis, defined by:
    \begin{equation}
        \label{eqn:r}
        r = e^{-\frac{1}{2}\left| z - {\bf y}^T Q {\bf y} \right|}
        \textrm{.}
    \end{equation}
    Here, $Q$ and $z$ are defined in the specification of the detector (Theorem \ref{thm:detector}).
\end{theorem}

\iflong
    \begin{IEEEproof}
        Assume ${\bf y}$ is given, having originated from the intruder model. Let $Q$ and $z$ be defined as in the presentation of the detector (Section \ref{sec:our-detector}), and consider the likelihood ratio of hypothesis 1 to hypothesis 2. With some effort, it can be shown from the intruder distributions (\ref{eqn:raw-detector}), that this ratio is given by:
        \begin{align}
            \frac{ \Pr \left( I = 1 \right) \Pr \left( {\bf Y} = {\bf y} \mid I = 1 \right) }
                 { \Pr \left( I = 2 \right) \Pr \left( {\bf Y} = {\bf y} \mid I = 2 \right) }
            = r_1
            = e^{-\frac{1}{2}\left({\bf y}^T Q {\bf y} - z\right)}
            \textrm{.}
        \end{align}
        Here, we restrict ${\bf y}$ to that satisfying ${\bf y}^T Q {\bf y} > z$, because $r_1$ is considered only when $\widehat{I} = 2$ (i.e. hypothesis 2 is more likely). For the case where $\widehat{I} = 1$, we find the opposite ratio, $r_2$, to be the following:
        \begin{equation}
            r_2
            = e^{-\frac{1}{2}\left(z - {\bf y}^T Q {\bf y}\right)}
            \textrm{,}
        \end{equation}
        which is similarly valid only for ${\bf y}^T Q {\bf y} \leq z$. Thus, we can write:
        \begin{equation}
            r = e^{-\frac{1}{2}\left|{\bf y}^T Q {\bf y} - z\right|}
            \textrm{,}
        \end{equation}
        where $r$ is interpreted as the likelihood ratio of the \textit{opposite} hypothesis, for all ${\bf y} \in \R^{k_f}$.
        
        To determine the probability of error given ${\bf y}$, consider the case where $I=1$ and $\widehat{I}=2$. We know from Bayes' theorem that
        \begin{equation}
            \Pr\left(I = 1 \mid {\bf Y} = {\bf y}\right)
            = \frac{\Pr\left(I = 1\right)}
                   {\Pr\left({\bf Y} = {\bf y}\right)}
            \Pr\left({\bf Y} = {\bf y} \mid I = 1\right)
            \textrm{.}
        \end{equation}
        Through simple manipulation, it can be shown that
        \begin{align}
            \nonumber
            \Pr \left( I = 1 \mid {\bf Y} = {\bf y} \right)
            &= \left( 1 + \frac{ \Pr \left( I = 2 \right) \Pr \left( {\bf Y} = {\bf y} \mid I = 2 \right) }
                               { \Pr \left( I = 1 \right) \Pr \left( {\bf Y} = {\bf y} \mid I = 1 \right) } \right)^{-1} \\
            \label{eqn:Pr-I1-given-yIh2}
            &= \left( 1 + \tfrac{1}{r_1} \right)^{-1}
            \textrm{.}
        \end{align}
        Similarly, for the case where $I=2$ and $\widehat{I}=1$, we find:
        \begin{equation}
            \label{eqn:Pr-I2-given-yIh1}
            \Pr \left( I = 2 \mid {\bf Y} = {\bf y} \right)
            = \left( 1 + \tfrac{1}{r_2} \right)^{-1}
            \textrm{.}
        \end{equation}
        As in the formulation of $r$, (\ref{eqn:Pr-I1-given-yIh2}) and (\ref{eqn:Pr-I2-given-yIh1}) are valid for the same complementary sets of ${\bf y}$, and hence the probability of detection error for all ${\bf y} \in \R^{k_f}$ can be written in terms of $r$:
        \begin{equation}
            \Pr \left( I \neq \widehat{I} \mid {\bf Y} = {\bf y} \right)
            = \left( 1 + \tfrac{1}{r} \right)^{-1}
        \end{equation}
    \end{IEEEproof}
\fi

\begin{table}[h!]
    \centering
    \begin{large}
    \begin{tabular}{c | c} 
        \hline
        Parameter & Value \\
        \hline\hline
        $k_1$     & 1    \\
        $k_2$     & 3    \\
        $m_1$     & 1    \\
        $m_2$     & 1    \\
        $  q$     & 1    \\
        $k_f$     & 20   \\
        $T$       & 0.5  \\
        $\Pr(I=1)$ & 0.5 \\
        $\Pr(I=2)$ & 0.5 \\ [1ex]
        \hline
    \end{tabular}
    \end{large}
    \caption{Table of simulation parameters.}
    \label{tab:simparams}
\end{table}

\section{Examples}
\label{sec:examples}

Simulations of the detector and error-probability calculation were undertaken for a case study, in which velocity measurement data of two intruder types (birds and drones) was synthesized according to \ref{eqn:intruder-model}. The parameters for the simulations are shown in Table \ref{tab:simparams}. 

The MAP detector has been implemented on the synthesized bird-drone data. Several results are shown. First, the implementation of the detector is illustrated, through comparison of the data test (left side of Equation 4) with the threshold $z$, for Monte-Carlo-generated trials of bird and drone data.

\begin{figure}[htbp]
    \includegraphics[width=\linewidth]{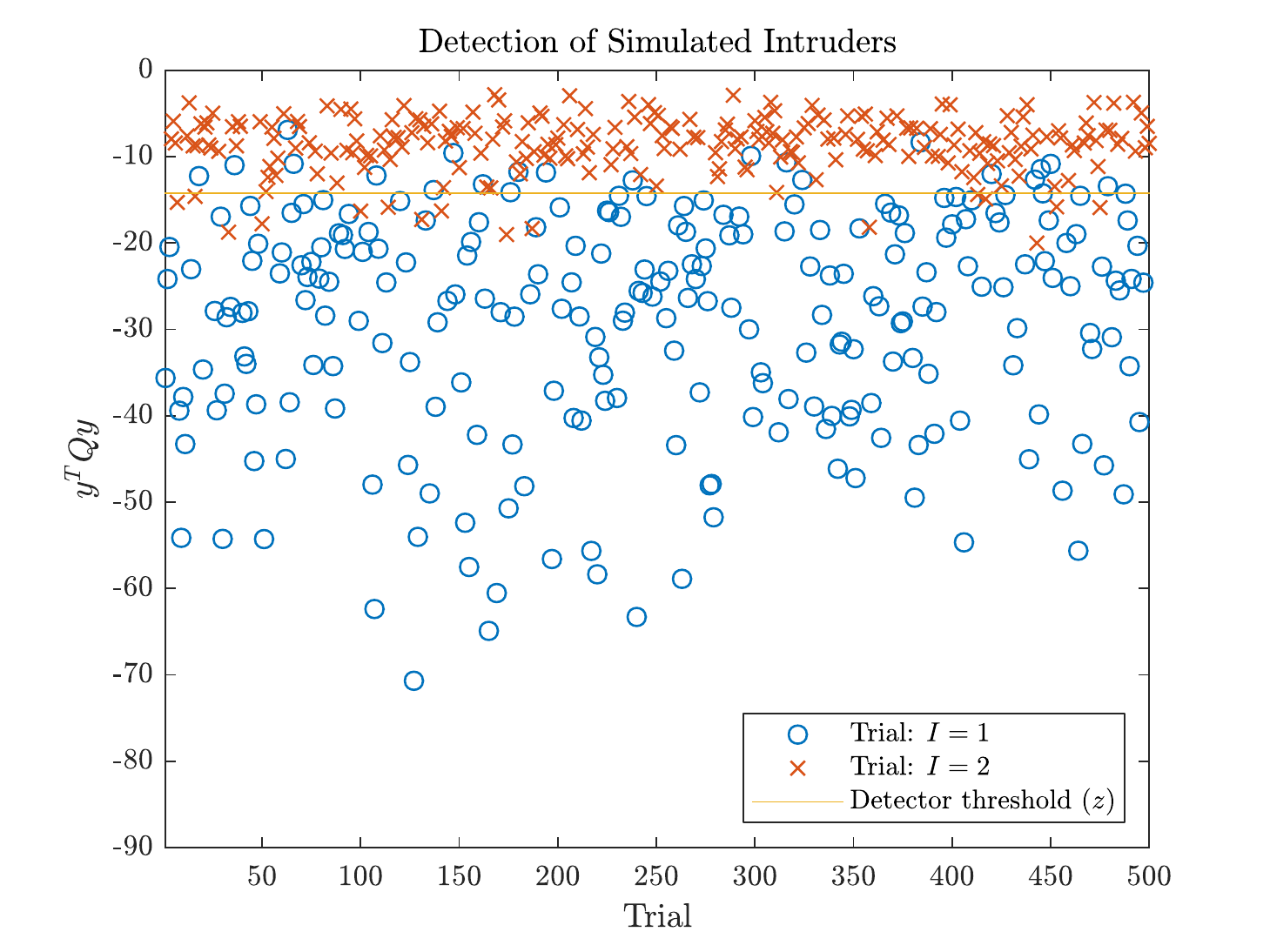}
    \caption{The implementation of the MAP detector is illustrated.  Specifically, the data test is
    compared with the detector threshold for trial bird and drone data, which have been synthesized
    according to the {\em a priori} probabilities and dynamic model.  The detector effectively
    distinguishes the two types of intruders.}
    \label{fig:500-trials}
\end{figure}

Second, real-time detection is illustrated, through application of the detector to streamed data over increasing horizons.  In this example, the detector converges to the correct hypothesis after a horizon of $k_f=12$.  It is noted that an example requiring a longer horizon for convergence is included for illustrative purposes: in most cases, convergence was achieved within $k_f=7$ samples.

\begin{figure}[htbp]
    \includegraphics[width=\linewidth]{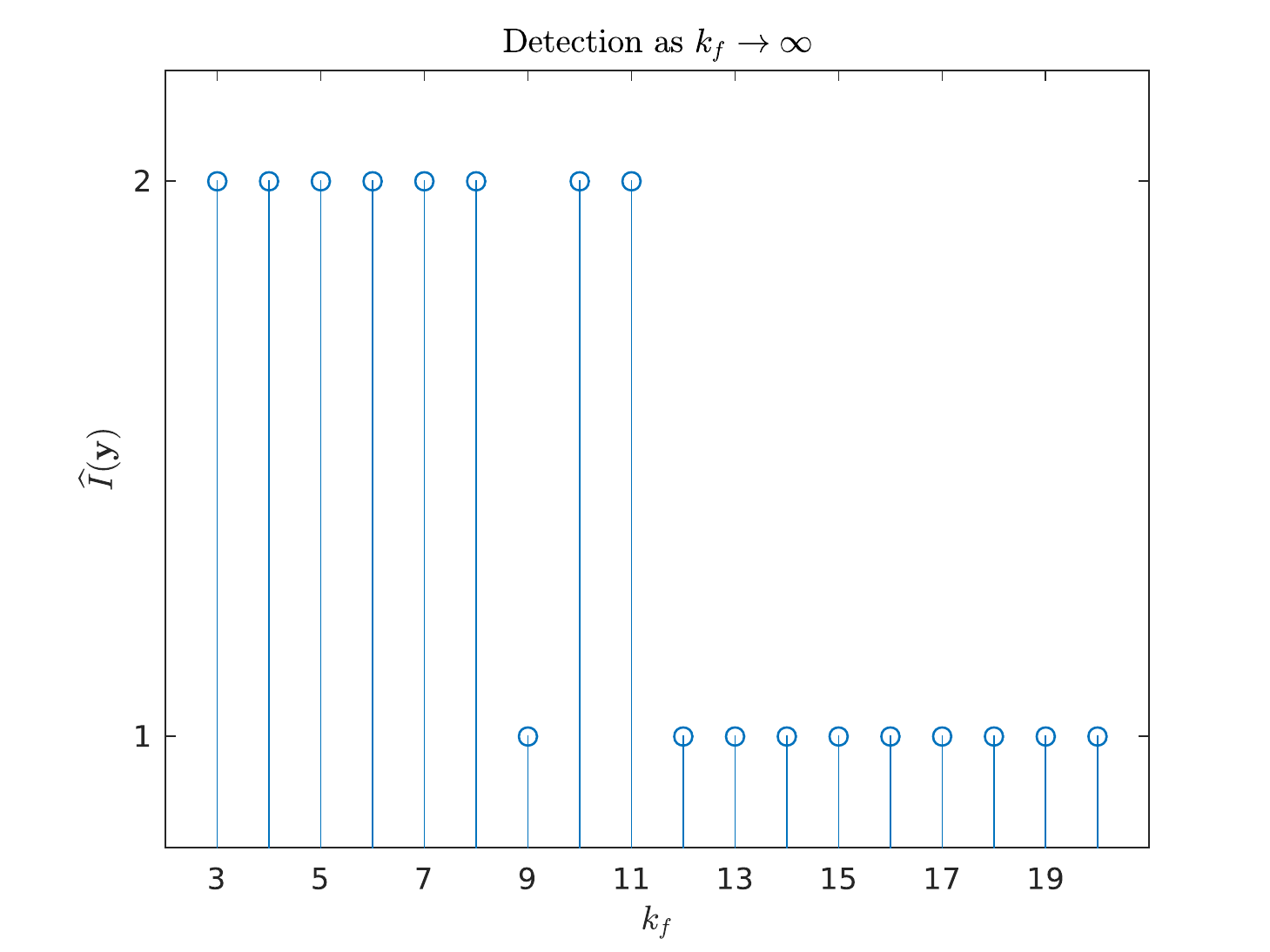}
    \caption{Application of the detector to streaming data is illustrated. The detector reaches correct hypothesis (intruder type) after $k_f=12$ samples.}
    \label{fig:detection-over-time}
\end{figure}

\begin{figure}
    \includegraphics[width=\linewidth]{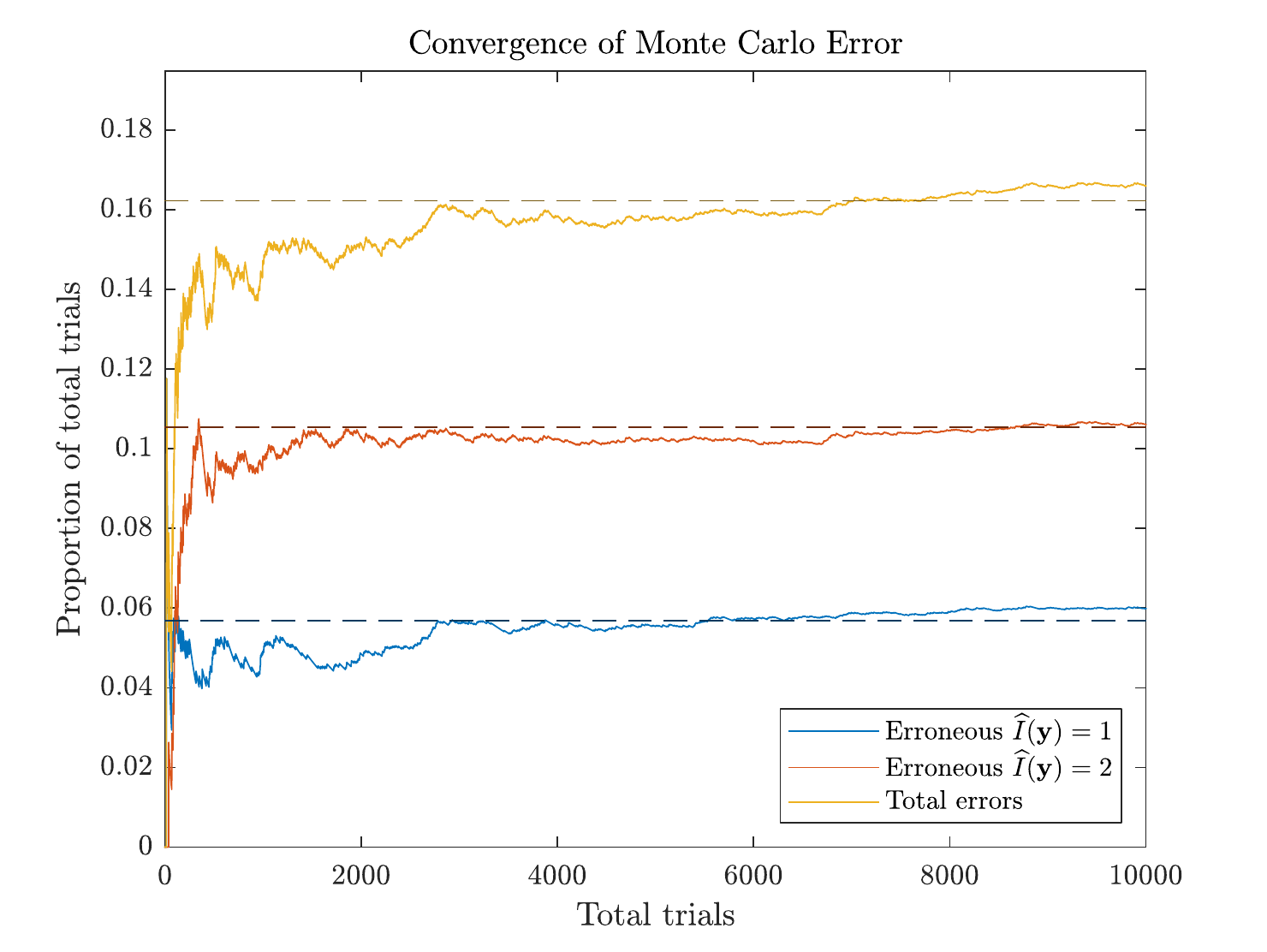}
    \caption{After several thousand trials, measured error is shown to converge to the computed
    error.}
    \label{fig:monte-carlo-comparison}
\end{figure}

Third, we compared the proportion of errors in repeated Monte Carlo trials to the expected total error as specified in Theorem 3, as a means to check the formal error calculation. The result after several thousand trials is shown in Figure \ref{fig:monte-carlo-comparison}: the total error probability as well as the error probability for each conclusion are shown.  The error probability as computed from Monte Carlo simulations over a long horizon matches the exact computation. 

Fourth, we illustrate the total probability of error as $k_1$, $m_1$ are held constant, and $k_2$, $m_2$ are varied, in order to get an idea for the feasibility of detection for a given intruder model pair. It can be seen that a (ratiometric) difference in $m$ improves detection fidelity more than that same difference in $k$. Additionally, it can be seen that when the time constant characterizing a drone ($m_2/k_2$) becomes significant relative to the sampling rate, it plays less of a role in detection than does a difference in the product of $k$ and $m$. This is intuitive, as aliasing will naturally play an integral role in the efficacy of the detector.

\begin{figure}[htbp]
    \includegraphics[width=\linewidth]{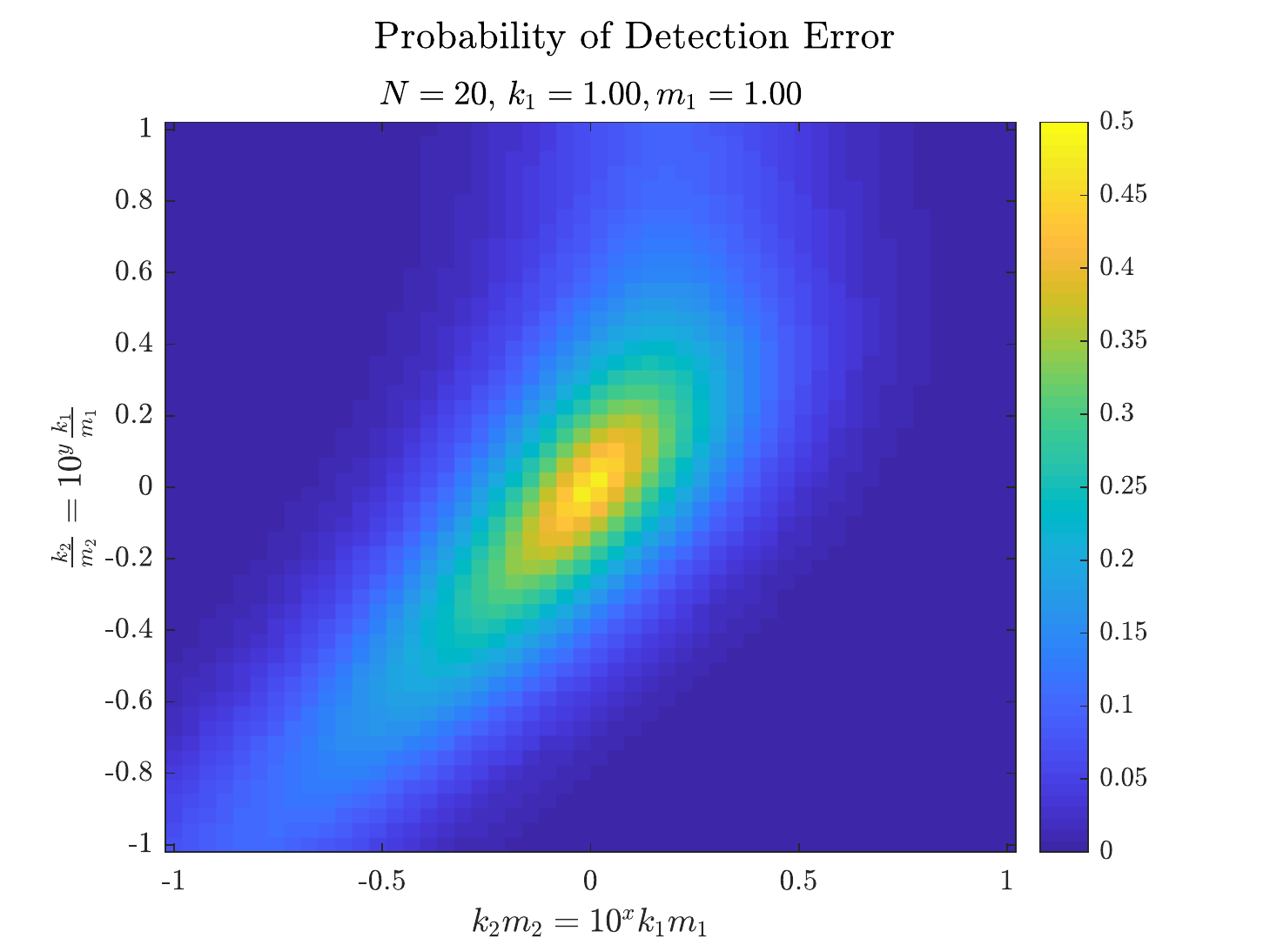}
    \caption{Detection error according to the relative ratios and products of $k$ and $m$, is illustrated on a log scale.}
    \label{fig:error-parameter-variation}
\end{figure}

Finally, we computed the detector's error probability as a function of $k_f$, to get an preliminary idea of how the detector's performance improves as the number of samples is increased.  The plot shows an exponential decrease in the error probability with the horizon $k_f$, which suggests that a modest number of points can be used for reliable detection.  In this sense, the detection approach appears to be promising for allowing robust differentiation of airspace intruders with different characteristics, such as drones and birds.

\bibliographystyle{IEEEtran}
\bibliography{main}

\begin{thebibliography}{1}
\providecommand{\url}[1]{#1}
\csname url@samestyle\endcsname
\providecommand{\newblock}{\relax}
\providecommand{\bibinfo}[2]{#2}
\providecommand{\BIBentrySTDinterwordspacing}{\spaceskip=0pt\relax}
\providecommand{\BIBentryALTinterwordstretchfactor}{4}
\providecommand{\BIBentryALTinterwordspacing}{\spaceskip=\fontdimen2\font plus
\BIBentryALTinterwordstretchfactor\fontdimen3\font minus
  \fontdimen4\font\relax}
\providecommand{\BIBforeignlanguage}[2]{{%
\expandafter\ifx\csname l@#1\endcsname\relax
\typeout{** WARNING: IEEEtran.bst: No hyphenation pattern has been}%
\typeout{** loaded for the language `#1'. Using the pattern for}%
\typeout{** the default language instead.}%
\else
\language=\csname l@#1\endcsname
\fi
#2}}
\providecommand{\BIBdecl}{\relax}
\BIBdecl

\bibitem{valvanis}
K.~Valvanis and G.~Vachtsevanos, Eds., \emph{Handbook of Unmanned Aerial
  Vehicles}.\hskip 1em plus 0.5em minus 0.4em\relax Dordrecht: Springer
  Netherlands, 2015, vol.~1.

\bibitem{uavsecurity}
T.~Humphreys, ``Statement on the security threat posed by unmanned aerial
  systems and possible countermeasures,'' 2015.

\bibitem{counteruas}
G.~Herrera, J.~Dechant, E.~Green, and E.~Klein, ``Technology trends in small
  unmanned aircraft systems (suas) and counter-uas: A five year outlook,'' no.
  IDA-P-8823, H-17-000624, 2017.

\bibitem{papoulis}
A.~Papoulis and S.~Pillai, \emph{Probability, Random Variables, and Stochastic
  Processes}.\hskip 1em plus 0.5em minus 0.4em\relax Tata McGraw-Hill
  Education, 2002.

\bibitem{standard-version}
D.~Petrizze, K.~Koorehdavoudi, M.~Xue, and S.~Roy, ``Distinguishing aerial
  intruders from trajectory data: A model-based hypothesis-testing approach,''
  \emph{American Control Conference}, 2021.

\bibitem{kms-inverse}
\BIBentryALTinterwordspacing
M.~Kac, W.~L. Murdock, and G.~Szegö, ``On the eigen-values of certain
  hermitian forms,'' \emph{Journal of Rational Mechanics and Analysis}, vol.~2,
  pp. 767--800, 1953. [Online]. Available:
  \url{http://www.jstor.org/stable/24900353}
\BIBentrySTDinterwordspacing

\bibitem{linear-models-chi-squared-distribution}
A.~C. Rencher and G.~B. Shaalje, \emph{Linear Models in Statistics}.\hskip 1em
  plus 0.5em minus 0.4em\relax John Wiley \& Sons, Ltd, 2007, ch.~5, pp.
  117--119.

\bibitem{davies-inverse}
\BIBentryALTinterwordspacing
R.~B. Davies, ``Numerical inversion of a characteristic function,''
  \emph{Biometrika}, vol.~60, no.~2, pp. 415--417, 1973. [Online]. Available:
  \url{http://www.jstor.org/stable/2334555}
\BIBentrySTDinterwordspacing

\end{thebibliography}

\end{document}